\documentclass[aps,prl,twocolumn,superscriptaddress,showpacs,floatfix,nofootinbib]{revtex4-1}
\usepackage{amsmath,graphicx,color}
\usepackage{hyperref}
\usepackage[x11names]{xcolor}
\usepackage{bm}
\usepackage{babel}
\usepackage{amsfonts,amssymb}

\def\TRENTo{{\sc t\kern-.05em \lower.5ex\hbox{r}\kern-.025em e\kern-.05em n\kern-.05em t\kern-.09em}o}
\def\iccing{{\sc i\kern-.05em c\kern-.05em c\kern-.05em i\kern-.05em n\kern-.05em g\kern-.05em}}
\def\ccake{{\sc c\kern-.05em c\kern-.05em a\kern-.05em k\kern-.05em e\kern-.05em}}
\def\vUSPhydro{v-{\sc u\kern-.05em s\kern-.05em p\kern-.05em}hydro}
\graphicspath{ {figures/} }

\begin{document}

\title{Mesoscopic chemical potentials across the (hyper)nuclear landscape}

\author{Jacquelyn Noronha-Hostler}
\affiliation{The Grainger College of Engineering, Illinois Center for Advanced Studies of the Universe, Department of Physics, University of Illinois at Urbana-Champaign, Urbana, IL 61801, USA}

\date{\today}

\begin{abstract}
Finite nuclei constrain the dense-matter equation of state (EOS), yet they are self-bound quantum
droplets far from the thermodynamic limit. Motivated by an analogy to quantum dots, we show that
the nuclear chart nevertheless defines a \emph{mesoscopic} regime in which \emph{mesoscopic
chemical-potential analogs} $\{\mu_B,\mu_Q,\mu_S\}$ can be extracted directly from nuclear and
hypernuclear binding energies after consistent Coulomb subtraction. These are discrete
finite-difference response functions---local slopes of the strong-interaction energy landscape---not
equilibrium grand-canonical chemical potentials. The nuclear chart itself supplies
an ``ensemble of nearby droplets'': finite differences across neighboring nuclei suppress
shell- and pairing-scale oscillations while retaining the smooth bulk trend, producing robust
slopes without a macroscopic limit. Thus, the data provide empirical local derivatives that any
strangeness-enabled EOS must reproduce near saturation. Mapping the measured (hyper)nuclear
landscape at $T\simeq 0$, we find smooth, numerically stable responses, including a large,
negative strangeness chemical-potential analog, and we identify specific hypernuclear measurements
that can directly test and sharpen these EOS constraints.
\end{abstract}

\maketitle

\noindent
\emph{Introduction.}
The strong force described by quantum chromodynamics (QCD) is responsible for binding together nuclei, confinement of quarks and gluons within hadrons, and is fundamental for the equation of state (EOS) of neutron-star matter \cite{Baym:2017whm}.  
The dense-matter community has long understood that nuclei can provide constraints on the EOS at low temperatures ($T$) and near nuclear saturation densities $n_{\rm sat}\sim 0.16\,{\rm fm}^{-3}$, see e.g.\ \cite{Lattimer:2015nhk,MUSES:2023hyz,Machleidt:2024bwl,Chatziioannou:2024tjq,Tsang:2012se,Roca-Maza:2015eza,Lonardoni:2014bwa}.  
Low-energy nuclear experiments, at approximately $T\sim 0$ MeV, can determine the properties of nuclei such as their mass $M(A,Z,N_\Lambda)$ and electric charge radii $R_C(A,Z,N_\Lambda)$ for different values of mass number $A$, proton number $Z$, and hyperon ($\Lambda$) number $N_\Lambda$.  
The experimentally measured values of masses and radii can be used to constrain  effective field theories \cite{Lonardoni:2017hgs,Lynn:2019rdt} and other nuclear structure approaches e.g.~\cite{Brown:2013mga,Roca-Maza:2018ujj}, and determine saturation properties. Once constrained, these frameworks can be used to calculate the dense matter EOS in a system of infinite volume.

A major challenge for equations of state is that finite nuclei are most naturally treated within the canonical ensemble, where particle numbers $\{A,Z,N_\Lambda\}$ are exactly conserved, whereas the equation of state relevant for neutron stars is typically formulated in a grand-canonical framework in terms of chemical potentials. 
In QCD, however, the fundamental conserved quantities of the strong interaction are the net baryon number $B$, electric charge $Q$, and strangeness $S$, rather than the particle numbers $\{A,Z,N_\Lambda\}$. While $\{A,Z,N_\Lambda\}$ provide a convenient labeling of finite nuclei, they are therefore only proxies for the conserved charges $\{B,Q,S\}$, and the mapping between these variables is not generally invertible once additional hadronic degrees of freedom are allowed (see e.g.\ \cite{Dexheimer:2008ax,Bombaci:2016xzl,Cruz-Camacho:2024odu}). Because chemical potentials are the Lagrange multipliers enforcing conserved charges in a grand-canonical description, a thermodynamically consistent formulation of dense matter is naturally expressed in terms of $\{\mu_B,\mu_Q,\mu_S\}$ (with $\mu_S$ used when strangeness is treated as a strong-interaction charge). 

In this work, we argue that effective \emph{mesoscopic} chemical-potential analogs---defined as discrete finite-difference response functions extracted from nuclear and hypernuclear binding energies---can be obtained directly from nuclear data without requiring a global inversion of $\{A,Z,N_\Lambda\}$ into $\{B,Q,S\}$.
We propose a new method of using experimentally measured properties of nuclei as effective \emph{mesoscopic} response functions (energy changes under adding or subtracting a conserved charge) to extract chemical-potential analogs. 
While nuclei are (quantum) microscopic systems, sensitive to pairing and shell effects, they can be analogous to mesoscopic systems such as quantum dots \cite{Tarucha:2001ynd}, where effective chemical potentials are defined through energy differences across a sequence of finite systems rather than through a thermodynamic limit. Nuclei provide a particularly clean realization of this idea, as they are self-bound and labeled by the conserved charges $B$ and $Q$ (and, in the strong sector, $S$). 
We interpret the emergence of the smooth, stable chemical-potential analogs found here under such finite-difference averaging as reflecting the self-bound droplet nature of nuclei near saturation, where bulk trends can be separated from finite-size fluctuations by averaging across neighboring systems.

Our approach provides a mapping between commonly used quantities in nuclear astrophysics such as the charge fractions of electric charge $Y_Q=Q/B$, strangeness $Y_S=S/B$, and the isospin asymmetry $\delta_I=1+Y_S-2Y_Q$ \cite{Yang:2025wop,Danhoni:2025qpn}, and the corresponding effective \emph{mesoscopic} chemical-potential analogs 
in a $T\sim 0$, near-saturation regime of the QCD phase diagram. 
We rely almost exclusively on experimental data and only use a minimal amount of modeling for the electric charge radii when $M(A,Z,N_\Lambda)$ has been measured but $R_C(A,Z,N_\Lambda)$ is not available. 
Uncertainties are quantified through standard error propagation (in the uncorrelated approximation). 
Our results provide stringent constraints on EOS models capable of reproducing saturation properties and provide a connection between canonical and grand-canonical descriptions. 

Throughout this paper we will use natural units of $\hbar=c=1$.

\medskip

\noindent
\emph{Methodology.}
Our effective mesoscopic chemical-potential analogs are calculated by looking at the change in strong force energy $\tilde{E}$ when adding/subtracting a conserved quantity, while holding the other quantum numbers fixed. 
At vanishing $T$ and for self-bound, ground-state nuclei (zero external pressure), variations of the strong-force energy reduce to changes in conserved quantum numbers of nuclei, the proton number $Z$, mass number $A$, and number of $\Lambda$ baryons
\begin{eqnarray}\label{eqn:totalderivativeALL}
\mathrm{d}\tilde{E}
&=&
\left(\frac{\partial \tilde{E}}{\partial A}\right)_{Z,N_\Lambda,V,\tilde{S}}\,\mathrm{d}A
+
\left(\frac{\partial \tilde{E}}{\partial Z}\right)_{A,N_\Lambda,V,\tilde{S}}\,\mathrm{d}Z\nonumber\\
&+&
\left(\frac{\partial \tilde{E}}{\partial N_\Lambda}\right)_{A,Z,V,\tilde{S}}\,\mathrm{d}N_\Lambda
.
\end{eqnarray}
where $V$ is the volume and $\tilde{S}$ is the entropy. 
Eq. (\ref{eqn:totalderivativeALL}) is written in nuclear labels $\{A,Z,N_\Lambda\}$; to rewrite it in terms of QCD conserved charges, we now express it in the $\{B,Q,S\}$ basis. For nuclei and $\Lambda$-hypernuclei in the restricted $\{n,p,\Lambda\}$ sector one has $B=A$, $Q=Z$, and $S=-N_\Lambda$.
The calculations of $\mu_B,\mu_S,\mu_Q$ are defined as:
\begin{equation}
    \mu_S=\frac{\partial \tilde{E}}{\partial S}\bigg|_{B,Q},\quad \mu_B=\frac{\partial \tilde{E}}{\partial B}\bigg|_{S,Q},\quad \mu_Q=\frac{\partial \tilde{E}}{\partial Q}\bigg|_{B,S},
\end{equation}
where we emphasize that in this work these derivatives are understood as \emph{mesoscopic} chemical potential analogs evaluated using controlled finite differences on the discrete nuclear chart.
Because nuclei form a discrete lattice in $\{A,Z,N_\Lambda\}$ space, chemical potentials are obtained from controlled finite-difference response functions, analogous to effective chemical potentials in quantum dots that are defined through energy differences between neighboring charge states, rather than through a thermodynamic limit \cite{Tarucha:2001ynd}.

In our calculations, we use finite-difference midpoint formulas on this discrete $\{A,Z,N_\Lambda\}$ space where the step size is fixed to $h=1$,  for order $\mathcal{O}(h)$ (includes 2 mirror (hyper)nuclei), $\mathcal{O}(h^2)$ (includes 3  mirror (hyper)nuclei), and $\mathcal{O}(h^4)$ (includes 5  mirror (hyper)nuclei). For $\mu_S$ we only have 3 possible states: $S=0,-1,-2$, so only up to $\mathcal{O}(h^2)$ is possible.  
Using $\mu_S$ as an example, the  $\mathcal{O}(h)$ (Euler) method is:
\begin{equation}
\mu_S^E\bigl(A,Z,\tfrac{N_{\Lambda,1}+N_{\Lambda,2}}{2}\bigr)=-\frac{\tilde{E}_2\left(A,Z,N_{\Lambda,2}\right)-\tilde{E}_1\left(A,Z,N_{\Lambda,1}\right)}{N_{\Lambda,1}-N_{\Lambda,2}}
\end{equation}
where the tilde signifies only the strong-force contribution to the total energy, and the two mirror (hyper)nuclei are 1 and 2. 
The $\mu_S$ picks up a negative sign because a strange quark provides a $S=-1$ contribution (i.e $S=-N_\Lambda$). 
For the accuracy to $\mathcal{O}(h^2)$ (mid-point) method we require knowledge of  $S=0,-1,-2$ states at a fixed $A,Z$. 
We first calculate $S=-1/2$ and $S=-3/2$ using the Eulerian method and use those values to calculate the midpoint i.e. 
\begin{eqnarray}
\mu_S^{m}\left(A,Z,-1\right)&=&\frac{\mu_S^E\left(A,Z,-\frac{1}{2}\right)+\mu_S^E\left(A,Z,-\frac{3}{2}\right)}{2}
\end{eqnarray}
where there is both an advantage in numerical accuracy but we also obtain results for integer values of $S=-1$ that can be aligned with specific hypernuclei. 
Most $S=-1$ results are from the midpoint method, but there are two exceptions for $A=11,Z=4$ and $A=12,Z=4$, since  data exists for double hypernuclei $^{12}_{\Lambda\Lambda}\rm{Be}$  and $^{11}_{\Lambda\Lambda}\rm{Be}$ \cite{E373KEK-PS:2013dfg} but not single hypernuclei data for $^{12}_{\Lambda}\rm{Be}$ and $^{11}_{\Lambda}\rm{Be}$ such that the Euler method (with a step size of $h=2$) allows us to calculate $S=-1$. 
In the Supplemental Material, we show the same Euler and midpoint methods for $\mu_B,\mu_Q$ and we have also worked out the $\mathcal{O}(h^4)$ accuracy derivatives that converge with the $\mathcal{O}(h^2)$ results. 

\begin{figure}
    \centering
     \includegraphics[width=\linewidth]{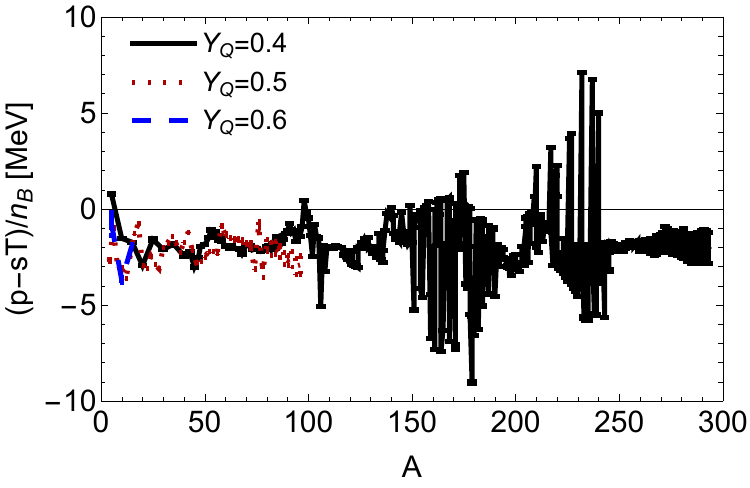}   
    \caption{We show the left-hand side of Eq.\ (\ref{eqn:Gibb_ps}), $(p-sT)/n_B$, plotted versus $A$ for different slices of $Y_Q$. We find values of approximately $\sim -3$ MeV (with error bars often consistent with zero) that do not appear to vary with $A$ or $Z$. This residual is much smaller than the extracted effective chemical potentials, so it represents a small correction relative to the extracted values.}
    \label{fig:pressure}
\end{figure}

Once the mesoscopic chemical potentials are known, we can check our assumption that we are both at vanishing temperatures (i.e. $T\rightarrow 0$) and that nuclei are self-bound (zero external pressure) using
\begin{equation}\label{eqn:Gibb_ps}
    \frac{p}{n_B}-\frac{s}{n_B}T=\mu_B+Y_Q\mu_Q+Y_S\mu_S-\frac{\varepsilon}{n_B}
\end{equation}
where the left-hand side should be small if the temperature is near zero,  we have accounted for all relevant EM corrections,  and our numerical results are accurate.
Eq.\ (\ref{eqn:Gibb_ps})  serves as a non-trivial self-consistency diagnostic of our assumptions that nuclei are self-bound and at vanishing temperature.

In Fig.\ \ref{fig:pressure} we show the result of the left-hand side of Eq.\ (\ref{eqn:Gibb_ps}) and found values of approximately $\sim -3$ MeV (with error bars that are often consistent with zero) and do not appear to vary with $A,Z$. 
This result supports that our systematic uncertainties are reasonably under control and that our assumptions are reliable. 
While understanding these systematics better would be interesting, they likely do not play a significant role in our extracted chemical potentials, which are 1-2 orders of magnitude larger than this systematic error.

\medskip

\noindent
\emph{Strong-force energy.} 
A crucial point to the calculations of the effective \emph{mesoscopic} $\{\mu_B,\mu_S,\mu_Q\}$ is that we must isolate the energy contributions coming just from the strong force. 
More specifically,  we want to determine the change in energy of the \emph{strong force} when we add either $\{A,Z,N_\Lambda\}$ to a nucleus. 
Assuming a nucleus is at rest, we only require the total mass 
\begin{equation}\label{eqn:mnucl_def}
    M(A,Z,N_\Lambda)=Zm_p+N_nm_n+N_\Lambda m_{\Lambda }-B_{\rm bind}(A,Z,N_\Lambda)
\end{equation}
where $m_p=938.2721$ MeV is the proton mass, $m_n=939.5654$ MeV is the neutron mass, and   $m_\Lambda=1115.683$ MeV  is the $\Lambda$ mass (we assume no errors since the error bars are orders of magnitude smaller than measured binding energies). 
Recall that the mass number within the nucleus is the sum of its constituents $A=Z+N_n+N_\Lambda$ (so $B=A$ in this restricted sector) such that the neutron number can always be calculated from the other particle numbers. 
The mass of the nucleus still contains unwanted Coulomb $E_C$ contributions that we must remove to access only the strong-force energy. 
The strong-force energy of a light nucleus is then
\begin{equation}
    \tilde{E}(A,Z,N_\Lambda)\equiv M(A,Z,N_\Lambda)-E_C(A,Z,N_\Lambda;R_C)
\end{equation}
where the Coulomb corrections depend on the charge radius of the nuclei $R_C(A,Z,N_\Lambda)$.
Here we include both direct and exchange Coulomb corrections within $E_C(A,Z,N_\Lambda;R_C)$ (see Supplemental Materials).

In this work we apply a data-driven approach where we use minimal theoretical input. 
For the masses of the light nuclei, we use the AME2020 tables, hypernuclei in \cite{Juric:1973zq,Pniewski:1985pc,1608.07448,JeffersonLabHallA:2018omd,FINUDA:2012yly,Wang:2021xhn,Hasegawa:1996fj}, heavy hypernuclei from \cite{Hasegawa:1996fj,Hotchi:2001rx}, and double hypernuclei in \cite{E373KEK-PS:2013dfg,Gal:2016boi} (see also \cite{Danysz:1963zza,Akaishi:1992pm,AGSE885:1999erv} for further discussion and possible measurements of $S=-2$ hypernuclei that were not included).
The $R_C$ data comes from the database in \cite{Angeli:2013epw}, that only covers approximately $25\%$ of the experimentally measured light nuclei.  The remaining light nuclei $R_C$ were estimated from phenomenological fits to the given experimental data. 
As of the time writing this paper, there was no available hypernuclei $R_C$ data.
Instead, we can estimate the change in the charge radius based on its core nucleus $\delta R=R_C(A,Z,N_\Lambda)- R_C(A-1,Z)$ using guidance from a  limited number of theoretical calculations   \cite{Tsushima:1997rd,Hiyama:2002yj,Hiyama:2009zz}.  
We provide tables of all our results (mesoscopic chemical potentials and $M,R_C,B_{\rm bind},E_C$) that can be downloaded here \cite{github}. 
\medskip

\noindent
\emph{Results.}
The  $\mathcal{O}(h)$ results lead to half-integer values of $\{B,Q,S\}$ such that one cannot calculate all 3 mesoscopic chemical potentials at the exact same point in $\{B,Q,S\}$ space. Only results with the $\mathcal{O}(h^2)$ produce integer values, making results directly comparable. 
When sufficient data is available, we prefer to show $\mathcal{O}(h^2)$ results, although it is not always possible for finite strangeness.  

\begin{figure}
    \centering
    \begin{tabular}{cc}
     \includegraphics[width=\linewidth]{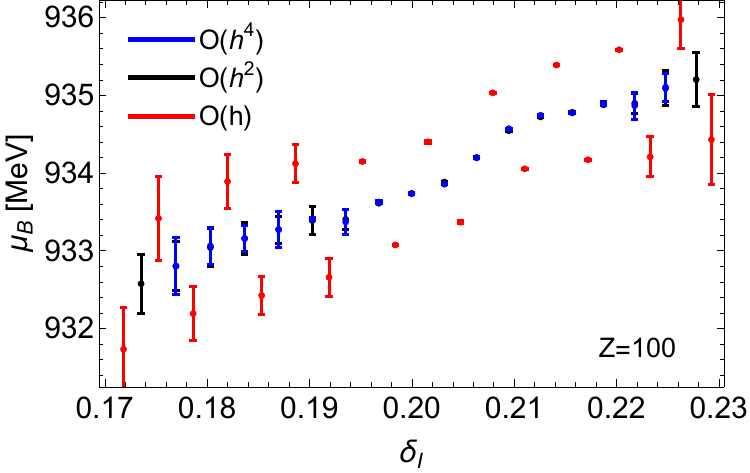}    
    \end{tabular}    
    \caption{Results for $\mu_B(A,Y_Q)$ vs isospin asymmetry. We compare a specific chain of isotopes at fixed $Z$ and their corresponding error bars. Calculations are performed at different orders of accuracy, using the Euler method $\mathcal{O}(h)$, mid-point method $\mathcal{O}(h^2)$, and $\mathcal{O}(h^4)$. }
    \label{fig:DERVcomp}
\end{figure}

In Fig.\ \ref{fig:DERVcomp} we compare our numerical results for $\mathcal{O}(h)$, $\mathcal{O}(h^2)$, and $\mathcal{O}(h^4)$ for $\mu_B(\delta_I)$ at fixed $Z=100$. 
We specifically choose a large $Z$ to illustrate the heavy-mass region. 
There $\mu_B(\delta_I)$ is directly correlated with $\delta_I$, where $\mu_B$ grows with $\delta_I$.
Our  $\mathcal{O}(h)$ method leads to fluctuations as one adds a neutron to the system, which arises from expected nuclear structure effects such as pair breaking/forming and shell structure. 
This is because the Euler method is really a separation energy between ($A,Z$ and $A-1,Z$). 
However, $\mathcal{O}(h^2)$ and $\mathcal{O}(h^4)$ suppress these effects, such that our results suppress finite-size structure effects. 
While not shown here, we found nearly identical results for $\mathcal{O}(h^2)$ and $\mathcal{O}(h^4)$ across multiple $A,Z$ values such that all further results will be shown only up to $\mathcal{O}(h^2)$.
The implication here is that discrete nuclear mass data, when processed with controlled finite-difference methods up to $\mathcal{O}(h^2)$ or higher, recover smooth mesoscopic response behavior across isospin. 
\footnote{Certain nuclear structure effects at large $A$ can lead to a non-monotonic behavior of $\mu_Q(Z)$ at fixed $A$.
We find a peak in  $\mu_Q$ at $Z=83$ because it is approaching the magic number of $Z=82$ and we also find a minimum at $\mu_Q$ for stable, very heavy nuclei.  We have checked that these effects are robust under both  $\mathcal{O}(h^2)$ and $\mathcal{O}(h^4)$, see the Supplemental Material.}

\begin{figure}
    \centering
    \includegraphics[width=\linewidth]{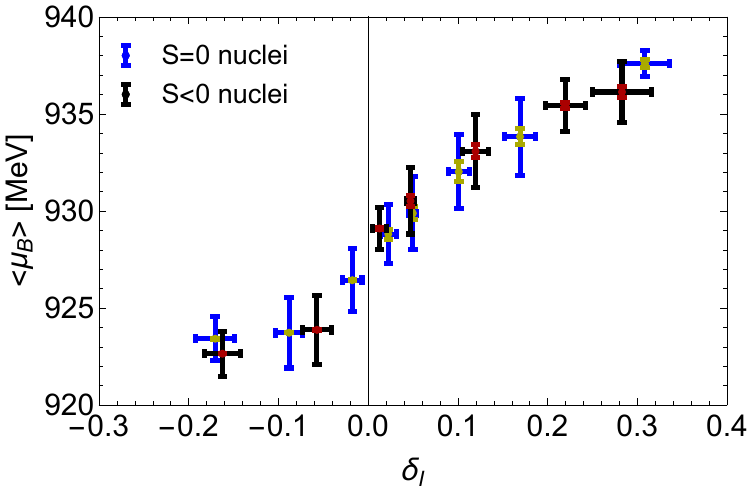}\\
    \includegraphics[width=\linewidth]{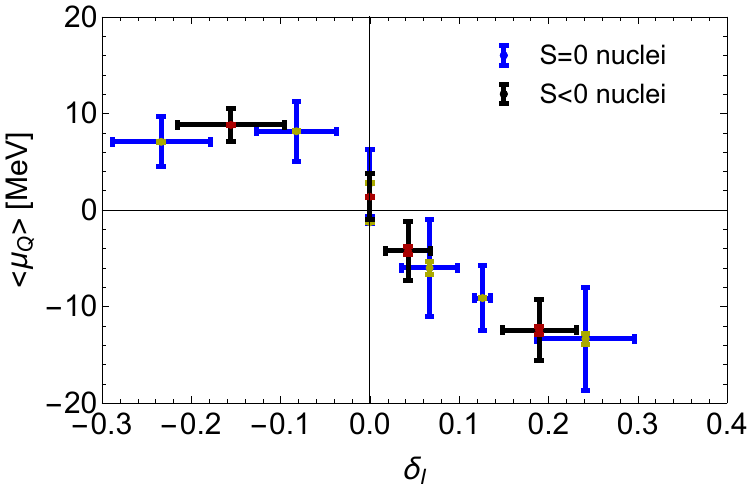}\\
    \includegraphics[width=\linewidth]{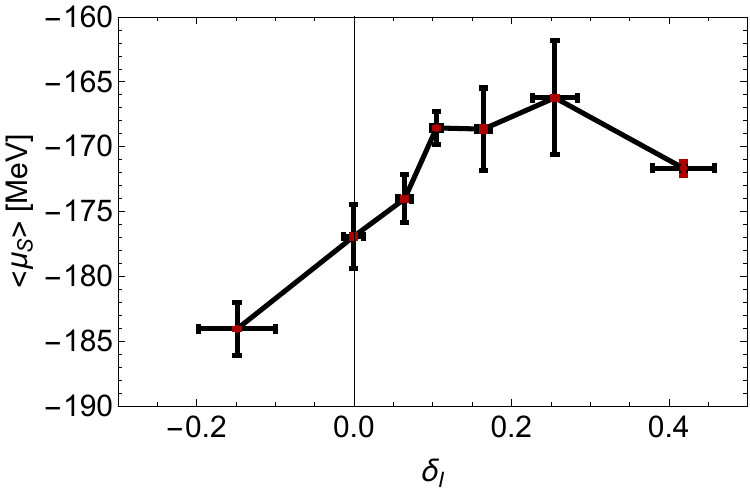}
    \caption{Averaged mesoscopic chemical potentials (top: baryon, middle: electric charge, bottom: strangeness), binned across the isospin asymmetry term. Statistical errors are shown in black/blue and systematic errors are in red/yellow.  }
    \label{fig:avg_mu}
\end{figure}

Our results predominantly scale with the isospin asymmetry, not with $A$.  Thus, we choose to plot averaged quantities:
\begin{equation}
    \langle \mu_X\rangle(\delta_I^{min};\delta_I^{max}) \equiv \frac{\sum_{\delta_I^{min}}^{\delta_I^{max}}A^{1/3}\mu_X}{\sum_{\delta_I^{min}}^{\delta_I^{max}}A^{1/3}}
\end{equation}
that are binned in ranges of $\delta_I$, where $X=B,Q,S$. Here we calculate statistical errors over the number of data points in a given bin size using jackknife statistics and we also propagate systematic errors using uncertainty quantification arising from experimental errors. 

In Fig.\ \ref{fig:avg_mu} we show our averaged mesoscopic chemical potentials across $\delta_I$.  We are dominated by statistical errors when averaging over our results. As more mass data on (hyper)nuclei is produced, these errors will lower. However, the propagated errors from the experiments themselves are significantly smaller. 
The range of negative $\delta_I$ is heavily limited by the proton drip line that occurs around $\delta_I\gtrsim-1/3$. 
Looking at the $\langle \mu_B\rangle (\delta_I)$ scaling, we find exactly what one would expect. Large isospin asymmetry (in the neutron-rich direction) leads to large chemical potentials because it is more energetically favorable to produce such nuclei (especially for heavy nuclei).  For isospin symmetric matter $\langle \mu_B\rangle(0)\sim 927$ MeV which is close to what many infinite nuclear matter calculations predict ($\mu_B^{V=\infty}\sim 922$ MeV). We find that $\langle \mu_B\rangle(\delta_I)$ is nearly identical for light nuclei and hypernuclei. 
The $\langle \mu_Q\rangle(\delta_I)$ corresponds perfectly to what one would expect from isospin asymmetry, it vanishes for isospin symmetry matter, is positive for $\delta_I<0$ and negative for $\delta_I>0$. Once again we find a strong consistency for light nuclei and hypernuclei results. 

Finally, we come to our $\langle \mu_S\rangle(\delta_I)$ results that demonstrate a large, negative $\langle \mu_S\rangle$.  The scaling of $\langle \mu_S\rangle(\delta_I)$ indicates that the energetic incentive for strangeness binding in nuclei weakens as systems become more neutron-rich.
Additionally, our results indicate that these mesoscopic chemical-potential analogs are far from the equilibrium weak-interaction condition $\mu_S=0$. Furthermore, we find an inflection around $\delta_I\sim 0.2$  that predominately comes from the influence of heavy hypernuclei data. Given that most heavy hypernuclei data is only available around $\delta_I\sim 0.2$, it is hard to draw strong conclusions here - but future experimental results across a range of $\delta_I$ would be extremely useful. 


\medskip
\noindent
\emph{Summary and Outlook}.
In this work we extracted the baryon, electric-charge, and strangeness mesoscopic chemical-potential analogs
$\{\mu_B,\mu_Q,\mu_S\}$ of (hyper)nuclei directly from experimental masses and charge radii, after
subtracting Coulomb contributions. Treating finite nuclei as self-bound systems at $T\simeq 0$ and employing
controlled finite-difference stencils on the nuclear chart, we showed that discrete nuclear data yield numerically
stable and convergent response functions. Experimental nuclei probe a broad and nontrivial region of mesoscopic
chemical-potential space, spanning $\mu_B \simeq 920$--$940~\mathrm{MeV}$,
$\mu_Q \simeq +10$ to $-15~\mathrm{MeV}$, and
$\mu_S \simeq -185$ to $-165~\mathrm{MeV}$,
across isospin asymmetries up to $|\delta_I|\lesssim 0.4$.

The key observation is that shell structure and pairing effects, while prominent nucleus by nucleus, sit atop a
smooth strong-interaction energy landscape controlled by nuclear saturation. By taking midpoint and higher-order
finite differences across neighboring nuclei—varying one conserved charge at a time while holding the others
fixed—these leading finite-size oscillations are suppressed, allowing robust local slopes of the energy landscape
to be extracted without invoking a thermodynamic limit.

A central result is the emergence of a \emph{large, negative strangeness chemical-potential analog} even at nearly vanishing net strangeness. In isospin-symmetric systems we find $\mu_Q\to 0$ within uncertainties, while $\mu_S$ remains sizable and negative, consistent with the statement that at fixed $A$ and $Z$ the strong-force energy $\tilde{E}$ increases as $S$ becomes more negative. This identifies a mesoscopic regime for finite, self-bound nuclei that is qualitatively distinct from dense matter in electroweak equilibrium, where strangeness is not conserved and $\mu_S=0$~\cite{Glendenning:1984jr}. Our analysis provides a direct mapping between commonly used nuclear/astrophysical variables—such as the isospin asymmetry $\delta_I$, electric-charge fraction $Y_Q$, and strangeness fraction $Y_S$—and the corresponding mesoscopic chemical-potential analogs $\{\mu_B,\mu_Q,\mu_S\}$.

Because these quantities are local slopes of the strong-interaction energy landscape, they offer a model-agnostic \emph{calibration and validation target} for macroscopic EOS that include strangeness, which can used to compare to ab initio approaches for hypernuclei e.g.~\cite{Hildenbrand:2024ypw,Tong:2025fzv,Haidenbauer:2025zrr,Wirth:2017bpw}. Equivalently, any strangeness-enabled EOS can be tested by comparing its ${\mu_B,\mu_Q,\mu_S}$—evaluated at $T\simeq 0$ and $n_B\simeq n_{\rm sat}$ under small variations of charge and strangeness around the same point in the phase diagram—to the empirical mesoscopic bands extracted here. This provides a direct, data-driven consistency check on the chemical-potential map that underlies hyperonic and strange-matter EOS modeling.

This framework also identifies targeted experimental measurements that would most strongly
sharpen these constraints. Improved binding-energy measurements of hypernuclei such as
$^{9}_{\Lambda}\mathrm{Be}$ or $^{8}_{\Lambda}\mathrm{Li}$ would allow the extraction of
$\{\mu_B,\mu_Q,\mu_S\}$ at fixed $\{A,Z,N_{\Lambda}\}$ for the first time. Data on
kaon-bound nuclei, such as $K^+$ or $K^0$ bound to $^{6}\mathrm{H}$,
$^{10}\mathrm{Be}$, or $^{12}\mathrm{B}$, would enable direct access to the $S \to 0$ limit of
$\mu_S$, providing a critical test of the large, negative strangeness chemical potential
inferred here. Proton-rich double-hypernuclei, for example
$^{10}_{\Lambda\Lambda}\mathrm{B}$, would further extend these studies into regimes of negative
isospin asymmetry at fixed strangeness.
Open-source databases such as \cite{Margueron:2025ugs} can play a vital role in enabling these calculations as new experimental data appear.

Ongoing and future experimental programs \cite{Chen:2025eeb}, including those at FAIR/GSI \cite{HypHI:2013sxa,Tanaka:2023onx,Saito:2023fnx}, together with emerging
analysis techniques for multi-strange systems \cite{He:2025vgg}, are well positioned to fill in these gaps in the
hypernuclear landscape. Such measurements would directly link laboratory nuclear data to mesoscopic constraints relevant to
dense, strange matter, strengthening the connection between nuclear
structure experiments and the equation of state relevant for neutron stars.

\emph{Acknowledgments} 
The author would like to thank Elizabeth Goldschmidt, Dean Lee, Veronica Dexheimer, Jorge Noronha, Mauricio Hippert, Yumu Yang for useful discussions on the implications of this work. 
We acknowledge
support from the US-DOE Nuclear Science Grant No. DE-SC0023861 and the National Science Foundation under the MUSES collaboration OAC2103680.

\bibliography{inspire,NOTinspire}

\onecolumngrid
\clearpage
\section{Supplemental Material}

\subsection{Derivation of the effective mesoscopic chemical-potential analog variables}

The total change of the strong-force contribution to the energy (as defined in the main text) can be written as
\begin{equation}
\mathrm{d}\tilde{E}
=
\left(\frac{\partial \tilde{E}}{\partial A}\right)_{Z,N_\Lambda,V,\tilde{S}}\,\mathrm{d}A
+
\left(\frac{\partial \tilde{E}}{\partial Z}\right)_{A,N_\Lambda,V,\tilde{S}}\,\mathrm{d}Z
+
\left(\frac{\partial \tilde{E}}{\partial N_\Lambda}\right)_{A,Z,V,\tilde{S}}\,\mathrm{d}N_\Lambda
+
\left(\frac{\partial \tilde{E}}{\partial V}\right)_{A,Z,N_\Lambda,\tilde{S}}\,\mathrm{d}V
+
\left(\frac{\partial \tilde{E}}{\partial \tilde{S}}\right)_{A,Z,N_\Lambda,V}\,\mathrm{d}\tilde{S}
.
\end{equation}
If one formally defines (for this energy function) the conjugate variables
\begin{equation}
    p=-\left(\frac{\partial \tilde{E}}{\partial V}\right)_{A,Z,N_\Lambda,\tilde{S}},
\end{equation}
and
\begin{equation}
    T=\left(\frac{\partial \tilde{E}}{\partial \tilde{S}}\right)_{A,Z,N_\Lambda,V},
\end{equation}
then the variations above take the familiar form used in the main text as a diagnostic/self-consistency relation.

The number of baryons in a nucleus, $A$, is directly equivalent to the baryon number for the systems considered here, such that
\begin{equation}
    \mu_B=\left(\frac{\partial \tilde{E}}{\partial A}\right)_{Z,N_\Lambda,V,\tilde{S}}.
\end{equation}
Given that for the experimentally measured (hyper)nuclei considered in this work the only electric-charge contribution comes from protons and the only strangeness contribution comes from $\Lambda$'s, we also have $Q=Z$ and $S=-N_\Lambda$, and therefore
\begin{eqnarray}
    \mu_Q&=&\left(\frac{\partial \tilde{E}}{\partial Z}\right)_{A,N_\Lambda,V,\tilde{S}},\\
    \mu_S&=&-\left(\frac{\partial \tilde{E}}{\partial N_\Lambda}\right)_{A,Z,V,\tilde{S}}\,.
\end{eqnarray}
Thus, the differential identity can be written as
\begin{equation}
\mathrm{d}\tilde{E}
=
\mu_B\,\mathrm{d}A
+
\mu_Q\,\mathrm{d}Z
-
\mu_S\,\mathrm{d}N_\Lambda
-
p\,\mathrm{d}V
+
T\,\mathrm{d}\tilde{S}
.
\end{equation}
In the main text, the quantities $\{\mu_B,\mu_Q,\mu_S\}$ are extracted as \emph{mesoscopic chemical-potential analogs} via controlled finite differences on the discrete nuclear chart, rather than as equilibrium grand-canonical chemical potentials in a thermodynamic limit.

What is then the consequence if $R$ does change? Let us assume a spherical nucleus:
\begin{eqnarray}
    V&=&\frac{4}{3}\pi R^3\\
    \mathrm{d}V&=&4\pi R^2\,\mathrm{d}R
\end{eqnarray}
where our total energy variation becomes:
\begin{equation}
\mathrm{d}\tilde{E}
=
\mu_B\,\mathrm{d}A
+
\mu_Q\,\mathrm{d}Z
-
\mu_S\,\mathrm{d}N_\Lambda
-
p\,4\pi R^2 \mathrm{d}R
+
T\,\mathrm{d}\tilde{S}
.
\end{equation}
Then, we still need the $\mathrm{d}R$ term:
\begin{equation}
    \mathrm{d}R=\left(\frac{\partial R}{\partial A}\right)_{Z,N_\Lambda,\tilde{S}}\,\mathrm{d}A
+
\left(\frac{\partial R}{\partial Z}\right)_{A,N_\Lambda,\tilde{S}}\,\mathrm{d}Z
+
\left(\frac{\partial R}{\partial N_\Lambda}\right)_{A,Z,\tilde{S}}\,\mathrm{d}N_\Lambda
+
\left(\frac{\partial R}{\partial \tilde{S}}\right)_{A,Z,N_\Lambda}\,\mathrm{d}\tilde{S}.
\end{equation}
Every time we take a total derivative with respect to a change in a particle number, it picks up a contribution from the $\mathrm{d}R$.
Thus,
\begin{equation}
    \frac{\mathrm{d}\tilde{E}}{\mathrm{d}A}\bigg|_{Z,N_\Lambda,\tilde{S}}=\mu_B -p\,4\pi R^2 \left(\frac{\partial R}{\partial A}\right)_{Z,N_\Lambda,\tilde{S}}
\end{equation}
\begin{equation}
    \frac{\mathrm{d}\tilde{E}}{\mathrm{d}Z}\bigg|_{A,N_\Lambda,\tilde{S}}=\mu_Q -p\,4\pi R^2 \left(\frac{\partial R}{\partial Z}\right)_{A,N_\Lambda,\tilde{S}}
\end{equation}
\begin{equation}
    \frac{\mathrm{d}\tilde{E}}{\mathrm{d}N_{\Lambda}}\bigg|_{A,Z,\tilde{S}}=-\mu_S -p\,4\pi R^2 \left(\frac{\partial R}{\partial N_{\Lambda}}\right)_{A,Z,\tilde{S}}
\end{equation}
where the pressure contribution is only significant if the nucleus is not self-bound (nonzero external pressure) and the radius varies significantly when adding/subtracting a particle.
Using the existing data on the charge radii (with the strong caveat that the charge radius is not equivalent to the total radius, but it at least allows us some way to estimate this effect), then we can at least calculate:
\begin{eqnarray}
    \delta R_Z(A,Z)&=&R_C(A,Z)-R_C(A,Z-1)\\
    \delta R_A(A,Z)&=&R_C(A,Z)-R_C(A-1,Z)
\end{eqnarray}
across all the experimentally measured charge radii. By averaging over all these differences, we find $\langle \delta R_A\rangle\sim 0.0076$ fm and $\langle \delta R_Z\rangle\sim 0.016$ fm. Since the radii are on average a few fm (let's assume $5$ fm for our estimate here), we can provide a back-of-the-envelope estimate of this contribution to $\mu_B,\mu_Q$.

For an order-of-magnitude bound, we note that the liquid--gas critical pressure is $\sim 0.3\,\mathrm{MeV/fm^3}$ at $T\sim 18$ MeV \cite{Elliott:2013pna}, so at the near-zero temperatures relevant for ground-state nuclei one expects any associated pressure scale to be much smaller than this. Taking $p\ll 0.3\,\mathrm{MeV/fm^3}$ as a conservative bound, we find
\begin{eqnarray}
    p\,4\pi R^2 \left(\frac{\partial R}{\partial A}\right)_{Z,N_\Lambda,\tilde{S}} &\ll& 0.7\, \mathrm{MeV} \\
    p\,4\pi R^2 \left(\frac{\partial R}{\partial Z}\right)_{A,N_\Lambda,\tilde{S}} &\ll& 1.5\, \mathrm{MeV}.
\end{eqnarray}
Thus, if the charge radii can at least provide reasonable estimates for $\delta R$, then the contribution from volume-change effects should be significantly smaller than our extracted mesoscopic chemical-potential analogs in this paper.

\subsection{Mesoscopic chemical-potential analogs up to $\mathcal{O}(h^4)$ accuracy}

The Euler and midpoint methods for the electric-charge mesoscopic chemical-potential analogs are:
\begin{eqnarray}
    \mu_Q^{E}\bigl( A,\frac{Z_1+Z_2}{2},N_{\Lambda}\bigr)&=&\frac{\tilde{E}_2\left(A,Z_2,N_{\Lambda}\right)-\tilde{E}_1\left(A,Z_1,N_{\Lambda}\right)}{Z_2-Z_1}\nonumber\\
    \end{eqnarray}
\begin{equation}
\begin{split}
\mu_Q^{m}\!\left(A,Z_2,N_\Lambda\right)
&= \frac{
\mu_Q^{E}\!\left(A,Z_2-\tfrac{1}{2},N_\Lambda\right)
+ \mu_Q^{E}\!\left(A,Z_2+\tfrac{1}{2},N_\Lambda\right)
}{2}\, .
\end{split}
\end{equation}
and analogously for the baryon mesoscopic chemical-potential analogs:
\begin{eqnarray}
    \mu_B^{E}\bigl( \frac{A_1+A_2}{2},Z,N_{\Lambda}\bigr)&=&\frac{\tilde{E}_2\left(A_2,Z,N_{\Lambda}\right)-\tilde{E}_1\left(A_1,Z,N_{\Lambda}\right)}{A_2-A_1}\nonumber\\
    \end{eqnarray}
\begin{equation}
\begin{split}
\mu_B^{m}\!\left(A_2,Z,N_\Lambda\right)
&= \frac{
\mu_B^{E}\!\left(A_2-\tfrac{1}{2},Z,N_\Lambda\right)
+ \mu_B^{E}\!\left(A_2+\tfrac{1}{2},Z,N_\Lambda\right)
}{2}\, .
\end{split}
\end{equation}

We can also work out the higher-order accuracy.
For light nuclei that have large chains of isotopes available, we can calculate derivatives with $\mathcal{O}(h^4)$ accuracy.
For these derivatives we require 5 isotopes, i.e.\ $A+2,A+1,A,A-1,A-2$ (or similar chains for $Z$), although the center value of $A$ does not contribute to the calculation.
Here we work out the example for the baryon mesoscopic chemical-potential analog, but analogous calculations were performed for $\mu_Q$.
With step size $h=1$ as used throughout this work, the standard five-point stencil gives
\begin{equation}
    \mu_B(A,Z)=\frac{-\tilde{E}(A+2,Z)+8\tilde{E}(A+1,Z)-8\tilde{E}(A-1,Z)+\tilde{E}(A-2,Z)}{12 }
\end{equation}
and the errors (assuming uncorrelated errors) are
\begin{equation}
    \sigma=\frac{1}{12}\sqrt{\sigma^2_{A+2}+64\sigma^2_{A+1}+64\sigma^2_{A-1}+\sigma^2_{A-2}}\, .
\end{equation}
In principle, one could also calculate forward or backward derivatives in the case of edges, but since we already find reasonable convergence comparing the midpoint method to our $\mathcal{O}(h^4)$ method, we do not perform those calculations here.

\subsection{Coulomb corrections}

Here we want to calculate the mesoscopic chemical-potential analogs related only to the strong force, such that we need to subtract any QED contributions (i.e.\ Coulomb corrections).
For nuclei, we will take into account two different sources that we must subtract from the total rest energy of a nucleus:
\begin{itemize}
    \item Direct Coulomb energy of a uniformly charged sphere:
    \begin{equation}
        E_C^{dir}=\frac{3}{5}\alpha \frac{Z(Z-1)}{R_C}
    \end{equation}
    \item Exchange Coulomb energy (Slater approximation) of a uniformly charged sphere
    \begin{equation}
        E_C^{ex}=-\frac{3}{4}\left(\frac{3}{2\pi}\right)^{2/3}\frac{\alpha}{R_C}Z^{4/3}
    \end{equation}
    \item Other corrections that may include: non-uniform charge distributions, radiation, magnetic moments, etc. Here we assume that they are subdominant and do not include these corrections.
\end{itemize}
Thus, our total Coulomb corrections are:
\begin{equation}
    E_C(Z,R_C)\equiv E_C^{dir}(Z,R_C)+E_C^{ex}(Z,R_C)
\end{equation}
where $R_C$ is the charge radius, which is not identical to the total nuclear radius $R$.
In experiments, the nuclear charge radius is what is typically available for nuclei whereas the total radius is significantly more difficult to measure (see e.g.\ \cite{Abrahamyan:2012gp,Horowitz:2012tj,PREX:2021umo,CREX:2022kgg,Giacalone:2023cet}).
Thus, in this work we use the database in \cite{Angeli:2013epw} that compiles $R_C$ extracted from low-energy nuclear experiments.

\subsubsection{Modeling $R_C$ when no experimental data exists}

\begin{figure}
    \centering
    \includegraphics[width=0.5\linewidth]{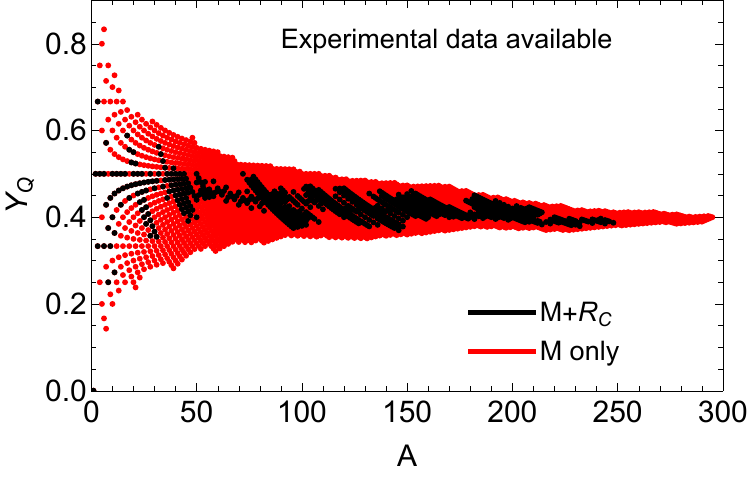}
    \caption{Scatter plot demonstrating the availability of experimental measurements of the mass of nuclei \cite{Huang:2021nwk,Wang:2021xhn} (red) versus when both the mass of nuclei and their corresponding charge radii are available (black).}
    \label{fig:availableRc}
\end{figure}

Unfortunately, not all nuclei have information about $R_C$ available from experimental data.
Using the database on the binding energies from \cite{Huang:2021nwk,Wang:2021xhn} we have 3558 nuclei whereas the compiled charge radii data exists only for 909 nuclei \cite{Angeli:2013epw}.
In Fig.\ \ref{fig:availableRc} we show a scatter plot of the available masses of nuclei (in red) compared to when there is both $R_C$ and $M$ information available (black).
We see that most of the $R_C$ information is available for nuclei close to isospin-symmetric matter and that some ranges of $A$ have significantly more available data than others. For instance, the heaviest nuclei above $A\gtrsim 250$ have no available data. There is also a sparser data set between $A=50$--$100$ and again around $A\sim 175$.
Overall, almost 75\% of our experimentally measured light nuclei are missing $R_C$ information.

In that case, we use the functional form:
\begin{eqnarray}\label{eqn:Rcfit}
    R_C(A,\delta_Q)\equiv r_0 A^{1/3}+r_1+r_2A^{-1/3}+r_3 \delta_Q+r_4\delta_Q^3
\end{eqnarray}
where $\delta_Q=(A-2Z)/A=1-2Y_Q$ is the electric charge isospin asymmetry (note this relation does not hold in the presence of strangeness \cite{Yang:2025wop,Danhoni:2025qpn} i.e. when $Y_S\neq 0$ then $\delta_I\neq \delta_Q$).
Here we performed a minimum $\chi^2$ fit over the known $R_C(A,\delta_Q)$ and their errors from \cite{Angeli:2013epw}. Doing so we obtained $\chi^2/d.o.f=0.132816$ and the coefficients: $r_0=0.928065$ fm, $r_1=0.0798712$ fm, $r_2=0.695799$ fm, $r_3=-0.550149$ fm, $r_4=-1.25474$ fm.

\begin{figure}
    \centering
    \begin{tabular}{cc}
      \includegraphics[width=0.5\linewidth]{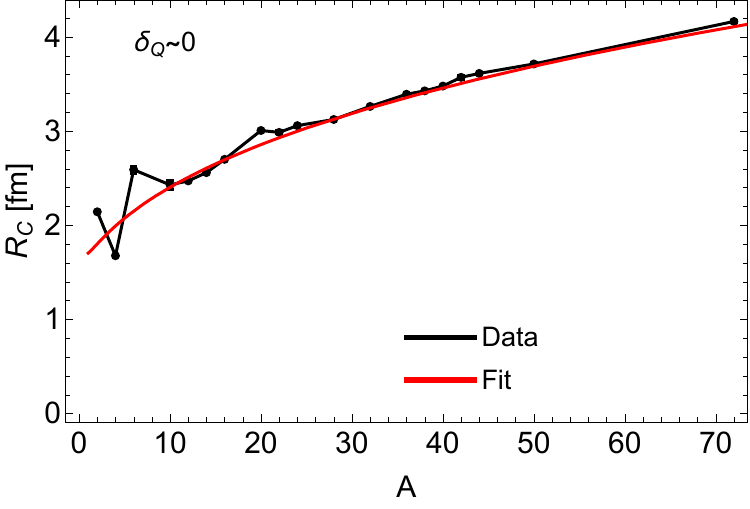}   &  \includegraphics[width=0.5\linewidth]{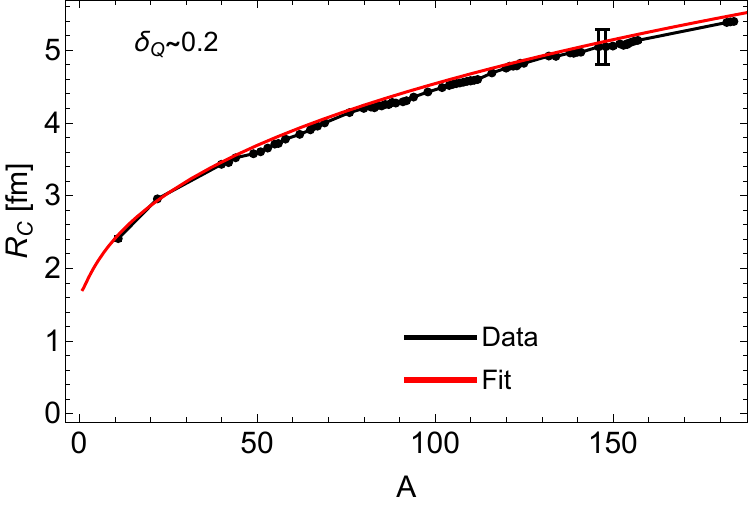}
    \end{tabular}
    \caption{Comparison of the experimentally measured nuclear charge radii from \cite{Angeli:2013epw} to our fit in Eq.\ (\ref{eqn:Rcfit}).}
    \label{fig:Rc_comp}
\end{figure}

When we have available $R_C$ from experimental data, we always use those values and propagate the error bars in our calculations of $E_C$.
However, when $R_C$ data is not available we assume a $10\%$ error on the value.
In Fig.\ \ref{fig:Rc_comp} we plot the charge radius vs the baryon number of the nuclei both for isospin-symmetric nuclear matter and isospin-asymmetric matter ($\delta_Q\sim 0.2$).
The experimental data is shown in black and the red lines are our fit. Generally, we find that our fit works quite well, especially for large $A$.
We also remind the reader that we always use the experimental data for the charge radii when possible, but only use the fit when data is not available.

\begin{figure}
    \centering
      \includegraphics[width=0.5\linewidth]{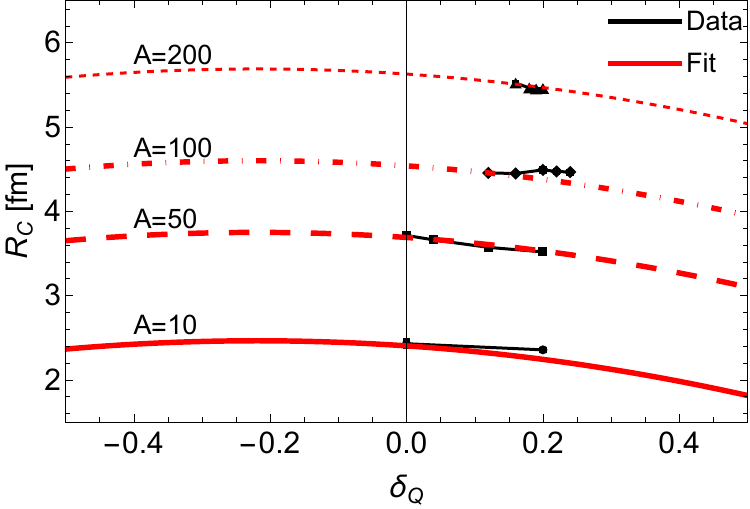}
    \caption{Comparison of the experimentally measured nuclear charge radii from \cite{Angeli:2013epw} to our fit in Eq.\ (\ref{eqn:Rcfit}). Here we show these results at fixed baryon number of the nuclei, plotted versus the charge-isospin asymmetry parameter $\delta_Q$.}
    \label{fig:Rc_DEL}
\end{figure}

In Fig.\ \ref{fig:Rc_DEL} we compare our results at fixed $A$ slices vs the isospin asymmetry parameter $\delta_Q$.
Once again, our fit does quite well compared to the data.
We note that significantly more data exist at fixed $\delta_Q$ when varying $A$, compared to the opposite of fixing $A$ and varying $\delta_Q$.
Thus, we do anticipate a bit more uncertainty for these fits if the $\delta_Q$ of a nucleus is far from the experimentally measured values of charge radii.

\subsubsection{Charge radii of hypernuclei}

A major challenge here is that at the time of writing this paper, we are not aware of experimental measurements of charge radii of hypernuclei.
Such measurements would be quite challenging since hypernuclei are unstable and very short lived.
Thus, we must take another approach to determine their charge radii.

One must turn to theoretical insights on the change of the charge radius of the core light nucleus $R_C(A-|S|,Z)$ compared to the radius of the hypernucleus itself, $R_C(A,Z,S)$.
While information on this change has been certainly been calculated (or extractable) in various theoretical frameworks, we are not aware of a single database or review paper that systematically studies this change across many hypernuclei, even in theoretical literature.
Instead, we only found a few papers \cite{Hiyama:2009zz,Tsushima:1997rd} that explicitly calculated a few differences between
\begin{equation}
    \delta R(A,Z,S)=\frac{R_C(A,Z,S)}{R_C(A-|S|,Z)}.
\end{equation}
For light hypernuclei, the example of $^{21}_\Lambda Ne$ ($R_C=3.8$ fm) vs $^{20}Ne$ ($R_C=4.0$ fm) was provided, which generally suggests a small decrease in the light nuclear core.
Another paper also implied this difference, see e.g.\ \cite{Hiyama:2002yj} but did not calculate the integrated $R_C$ directly.
For heavier hypernuclei in \cite{Tsushima:1997rd} it seems that the charge radius is nearly identical to the light core.
Although, an important caveat here is that this is primarily discussed for $\Lambda$ baryons that are electrically neutral; it is much harder to say how a charged hyperon would affect the nuclear structure.
Given that the experimental data we have at hand is entirely from $\Lambda$ hypernuclei, then we leave this challenge to future work.
Thus, here we assume that for hypernuclei of $A<28$, $\delta R(A,Z,S)\sim 0.95\pm 0.05$ (based on the example in \cite{Hiyama:2009zz} where $\delta R_C(A,Z,S)=0.95$) where we use our knowledge of the light nuclear core (previously discussed) to determine the hypernuclei charge radii.
For heavier hypernuclei of $A\geq 28$ we assume no change in the $R_C$.
Then for double hypernuclei we have even less information, so we assume an even larger error bar $\delta R_C(A,Z,S)\sim 0.95\pm 0.1$.
Note we once again use error propagation here such that these error bars enhance the original $R_C$ error from its corresponding light nuclear core.

\subsection{Calculation of nuclear masses}

While existing hypernuclei contain either a single $\Lambda$ ($S=-1$) or double $\Lambda\Lambda$ ($S=-2$), it is not impossible that other hyperons such as cascades ($S=-2$) or omegas ($S=-3$) could also exist. In particular, theoretical calculations of such hypernuclei do exist \cite{Gal:2016boi}.

Some experimental data used in this paper is not provided in terms of $M(A,Z)$ but rather in terms of the binding energy per nucleon $B_{\rm bind}(A,Z,S)/A$ (note that $(A,Z)\equiv (A,Z,S=0)$ for light nuclei). To obtain the mass of the nucleus, one must include all rest masses of the nucleons and hyperons inside i.e.
\begin{equation}\label{eqn:mnucl_def}
    M(A,Z,S)=Zm_p+(A-Z-|S|)m_n+|S|m_{\Lambda }-B_{\rm bind}(A,Z,S)
\end{equation}
where $m_p=938.2721$ MeV is the mass of the proton and $m_n=939.5654$ MeV is the mass of the neutron (we assume no errors since the error bars are orders of magnitude smaller than the binding energies discussed here). In the case for light nuclei the binding energy is provide \emph{per nucleon} such that one must multiply by $A$.
Furthermore, the hypernuclei binding energies $B_\Lambda(A,Z,S)$ for a single $\Lambda$ baryon are typically provided in respect to their ``core'' nucleus $B_{\rm bind}(A-1,Z)$ (the nucleus that would exist in absence of the $\Lambda$ particle) i.e.
\begin{eqnarray}
    B_\Lambda(A,Z,S)&\equiv&  B_{\rm bind}(A,Z,S)-B_{\rm bind}(A-1,Z)\\
    &=&M(A-1,Z)+m_\Lambda-M(A,Z,S)
\end{eqnarray}
where $B_{\rm bind}(A,Z,S)$ is actually the quantity that we need to calculate $M(A,Z,S)$ in Eq.\ (\ref{eqn:mnucl_def}).
For double hypernuclei, there are unfortunately two different definitions (total double--$\Lambda$ binding energy $B_{\Lambda\Lambda}(A,Z,S)$ vs $\Lambda\Lambda$ bond energy $\Delta B_{\Lambda\Lambda}$) used to report experimental data, which can provide some confusion.
The total double--$\Lambda$ binding energy $B_{\Lambda\Lambda}(A,Z,S)$ is the same principle as what we defined for $B_\Lambda$ in that you subtract the binding energy of the core nucleus i.e.
\begin{eqnarray}
    B_{\Lambda\Lambda}(A,Z,S)&\equiv&  B_{\rm bind}(A,Z,S)-B_{\rm bind}(A-2,Z)\\
    &=&M(A-2,Z)+2m_\Lambda-M(A,Z,S).
\end{eqnarray}
Alternatively, some experimental data is provided as the $\Lambda\Lambda$ bond energy
\begin{eqnarray}
   \Delta B_{\Lambda\Lambda}(A,Z,S)&\equiv&B_{\Lambda\Lambda}(A,Z,S) -2  B_{\Lambda}(A-1,Z,-1)\\
    &=&2M(A-1,Z,-1)-M(A-2,Z)-M(A,Z,S)
\end{eqnarray}
where $B_\Lambda$, $B_{\Lambda\Lambda}$, $\Delta B_{\Lambda\Lambda}$ are all the \emph{total} binding energies (i.e not per nucleon).

To summarize, for light nuclei we use the AME2020 tables \cite{Huang:2021nwk,Wang:2021xhn} that provide $B_{\rm bind}(A,Z,0)/A$ and must solve Eq.\ (\ref{eqn:mnucl_def}) to obtain $M(A,Z,0)$.
In this work, the experimental hypernuclear input involves only $\Lambda$ and $\Lambda\Lambda$ systems, so the relations above for $B_\Lambda$, $B_{\Lambda\Lambda}$, and $\Delta B_{\Lambda\Lambda}$ are sufficient to reconstruct $M(A,Z,S)$ from the reported binding-energy data.
For double $\Lambda$ hypernuclei when we are provided $\Delta B_{\Lambda\Lambda}$, we must solve:
\begin{equation}
    M(A,Z,S)=2M(A-1,Z,-1)-M(A-2,Z)-\Delta B_{\Lambda\Lambda}(A,Z,S).
\end{equation}

For this work, we have calculated tables of $M(A,Z,S)$ and $\tilde{E}(A,Z;R)$ and for the light nuclei in the AME2020 tables, hypernuclei in \cite{Juric:1973zq,Pniewski:1985pc,Botta:2016kqd,JeffersonLabHallA:2018omd,FINUDA:2012yly,Wang:2021xhn,STAR:2022zrf,ALICE:2022sco,ALICE:2024djx}, and double hypernuclei in \cite{E373KEK-PS:2013dfg,Gal:2016boi} (see also \cite{Danysz:1963zza,Akaishi:1992pm,AGSE885:1999erv} for further discussions).
We use error propagation assuming entirely uncorrelated errors (because we do not have a reasonable method to take correlations into account at this time).
Because hypernuclei do not appear to have standardized tables like the light nuclei, it makes it challenging to track down all available data and compare directly across experiments (especially when multiple data points exist). To help ensure reproducibility and adaptability as more data or improved methods arise, we have made these tables publicly available at \cite{github}.

We decided to stop our list at $A=28$, because heavier hypernuclei are so few and far between that they do not allow us to extract any meaningful information that relates to our generalized expansion. These heavy hypernuclei are all not for isospin symmetric matter and also only have a single state at their corresponding $A$ and $\delta_I$. We would require more states at a fixed point in either $\delta_I$, $A$, or $Y_S$ to be able to extract relevant information for the generalized symmetric expansion.
Thus, until further states are measured in these ranges, we do not include them in our table.
We also used \cite{Wang:2021xhn} to find light nuclei with similar $A$ and $\delta_I$ values for comparison.

\begin{figure}
    \centering
    \begin{tabular}{cc}
     \includegraphics[width=0.5\linewidth]{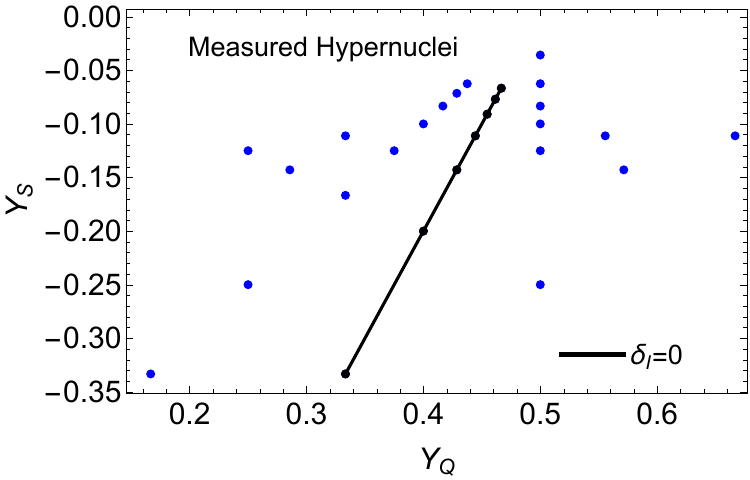}    & \includegraphics[width=0.5\linewidth]{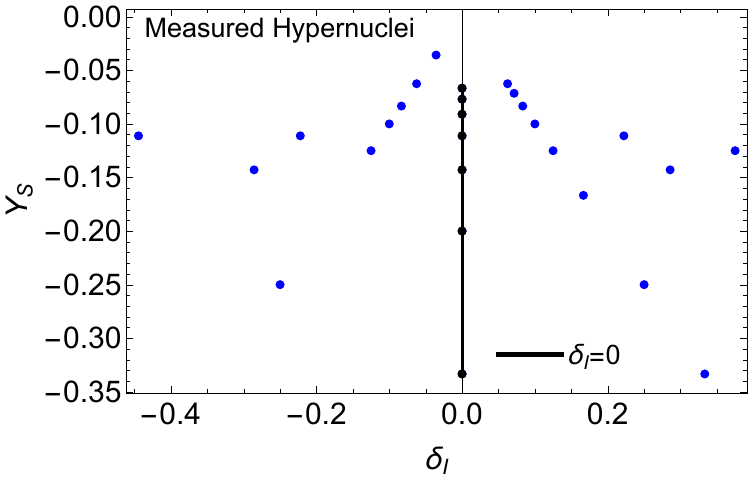}
    \end{tabular}

    \caption{Scatter plot of the measured hypernuclei for their strangeness fraction $Y_S=-N_\Lambda/A$ vs electric charge fraction $Y_Q=Z/A$ (left) and their $Y_S$ vs the isospin asymmetry $\delta_I=1+Y_S-2Y_Q$ (right). Isospin symmetric matter is highlighted in the solid black line.}
    \label{fig:measured_hyper}
\end{figure}

In Fig.\ \ref{fig:measured_hyper} we show a scatter plot of the available experimentally measured hypernuclei data for their binding energies. We see a relatively even distribution across isospin symmetric matter. Although, we remind the reader that when $Y_S\neq 0$ isospin symmetric matter does not correspond to $Y_Q=0.5$ but rather $Y_Q(\delta_I=0)=0.5+Y_S/2$. Thus, larger (negative) values of $Y_S$ decrease $Y_Q$, as we can see in the left plot in Fig.\ \ref{fig:measured_hyper}.

\subsection{Mesoscopic chemical-potential analogs $\mu_B,\mu_Q,\mu_S$ across the nuclear landscape}

\begin{figure*}
    \centering
    \begin{tabular}{ccc}
     \includegraphics[width=0.33\linewidth]{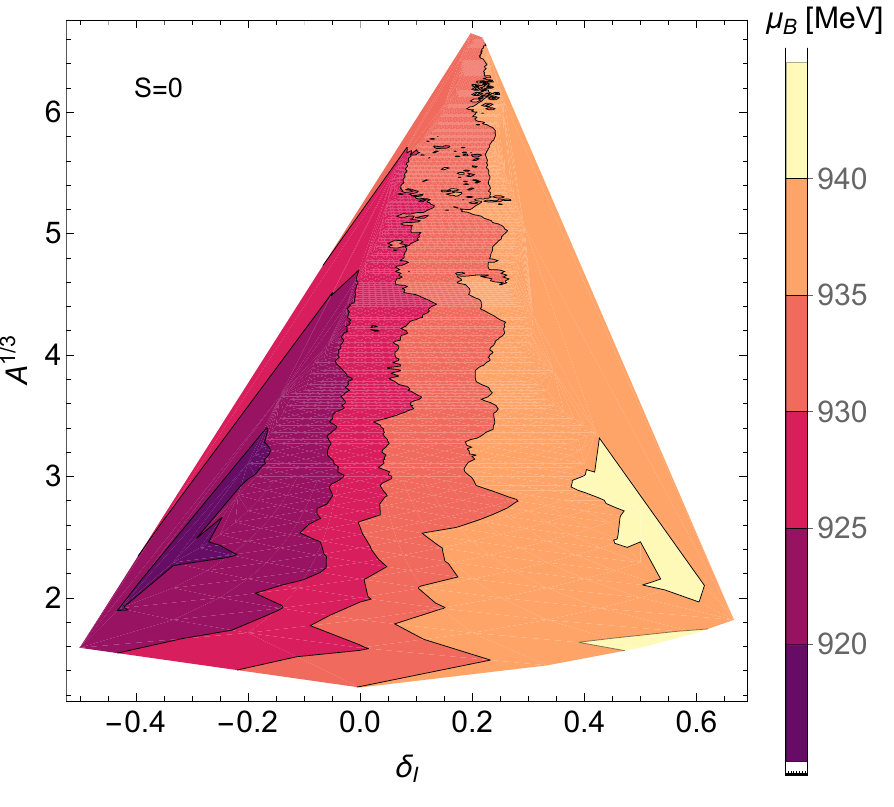}    & \includegraphics[width=0.33\linewidth]{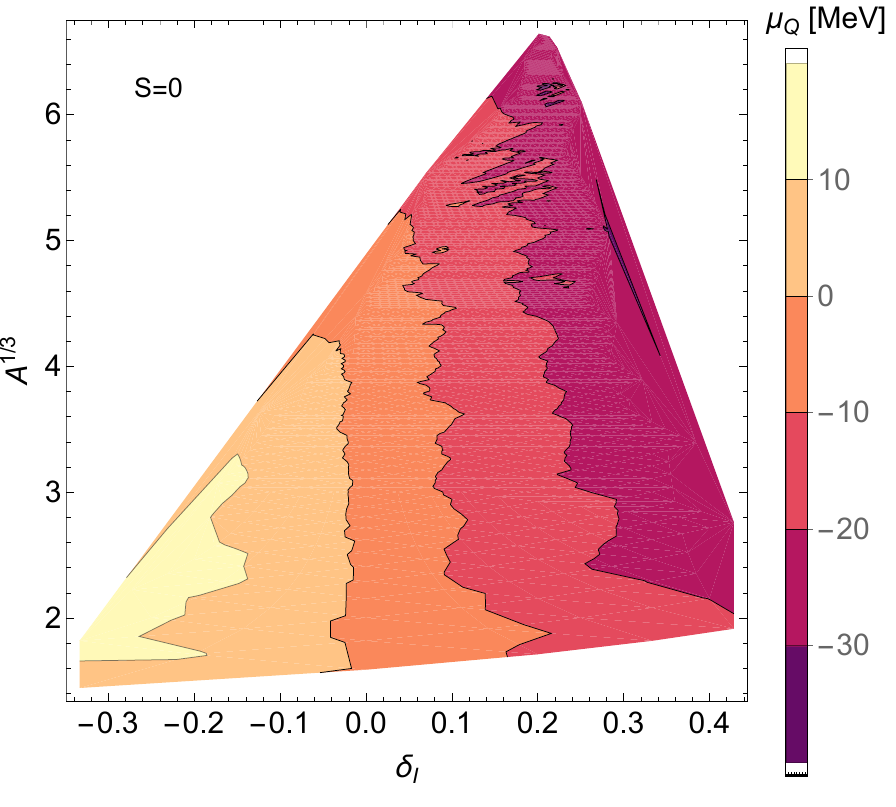} &
     \includegraphics[width=0.33\linewidth]{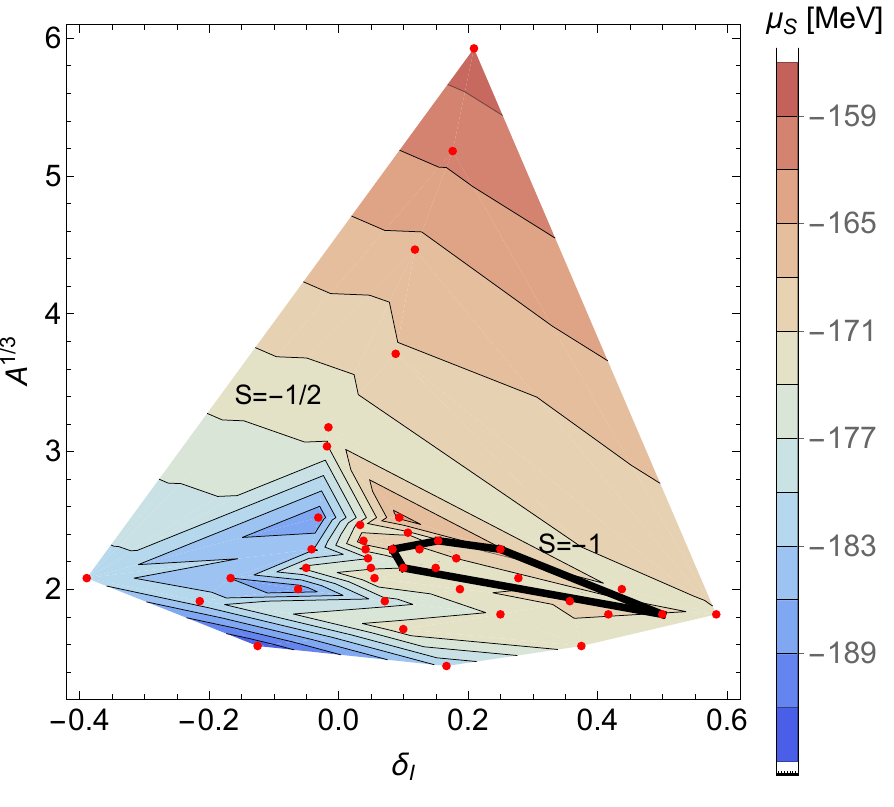}
    \end{tabular}
    \caption{Results for the mesoscopic chemical-potential analogs $\mu_B(A,Y_Q)$ (left), $\mu_Q(A,Y_Q)$ (middle), and $\mu_S(A,Y_Q)$ (right), plotted versus the isospin asymmetry $\delta_I$ and the expected volume scaling $A^{1/3}$. For the $\mu_S(A,Y_Q)$ results we plot the location of the existing hypernuclei data as red dots to highlight the scarcity of the data in the large-$A$ regime.}
    \label{fig:muBQS_results}
\end{figure*}

For the rest of the plots, we will only show the central values of the mesoscopic chemical-potential analogs calculated at $\mathcal{O}(h^2)$ to easily visualize their dependence on $\delta_I$ and the scaling with the volume $V\propto A^{1/3}$. However, readers should keep in mind both the propagated error and the systematic error on these results that cannot be easily visualized in such plots. Thus, errors on the order of a few MeV should be considered.
In Fig.\ \ref{fig:muBQS_results} we start with the light nuclei and plot $\mu_B^{m}(A,Z)$ (left) and $\mu_Q^{m}(A,Z)$ (middle) across the isospin asymmetry and scaling of the volume $A^{1/3}$.
We find that the $\mu_B$ values range from 918--940 MeV where proton-rich nuclei ($\delta_I<0$) reach smaller $\mu_B$ and neutron-rich nuclei ($\delta_I>0$) reach larger $\mu_B$. Around symmetric nuclear matter, we find values of $\mu_B\sim 930$ MeV, although for larger nuclei it decreases to $\sim 925$ MeV, approaching values one anticipates for infinite nuclear matter, i.e.\ $\mu_B\sim 922$ MeV.
We do not find a strong scaling with the volume, beyond a small shift to lower $\mu_B$ at fixed $\delta_I$ for larger volumes.
For $\mu_Q(A,Z)$ in Fig.\ \ref{fig:muBQS_results} (middle), at $\delta_I\sim 0$ we obtain the expected $\mu_Q\sim 0$ within uncertainties. Similarly, $\delta_I>0$ leads to negative $\mu_Q$ and $\delta_I<0$ leads to positive $\mu_Q$, which is expected in these specific isospin-asymmetric regimes. Heavy nuclei become more neutron-rich such that more data exists at positive $\delta_I$, which can reach up to $\mu_Q\sim -30$ MeV.
The jaggedness at large $A$ with $\delta_I\sim 0.2$--$0.3$ appears because of the approach of $Z\rightarrow 82$ combined with the minimum in $\mu_Q$ for stable nuclei.

We can estimate the order of magnitude for $\mu_S$ using $\mu_S=\Delta \tilde{E}/\Delta S$. Since we fix $A,Z$, as we add a $N_\Lambda$ we exchange a neutron for a $\Lambda$. Our dominant contribution to $\mu_S$ is the mass difference
\begin{equation}
\mu_S^{EST}\sim \frac{m_\Lambda-m_n}{-1}\sim -176~\mathrm{MeV}.
\end{equation}
Analogously, the expected mesoscopic chemical-potential analog for a $K^0$--nucleus bound system would be $\mu_S^{EST}\sim m_{K}/(\pm 1)\sim \pm 498$ MeV. Thus, the complexities of the strong force arise from variations around these expected values.
In Fig.\ \ref{fig:muBQS_results} (right) we plot $\mu_S$ across the isospin asymmetry and scaling with the volume.
The Euler method is shown for $S=-1/2$ results but overlaid on top are the $S=-1$ results outlined in black.
We find a large, negative $\mu_S$ that is nearly always more negative than $\mu_S^{EST}$, instead $\sim [-175,-195]$ MeV.
Proton-rich hypernuclei ($\delta_I<0$) lead to larger deviations from $\mu_S^{EST}$ whereas positive values $\delta_I>0$ have smaller magnitudes of $\mu_S$ such that $\mu_S\rightarrow \mu_S^{EST}$ for large, positive $\delta_I$.
We note that at large $A$ for hypernuclei the data is extremely limited.

\begin{figure}
    \centering
    \includegraphics[width=0.5\linewidth]{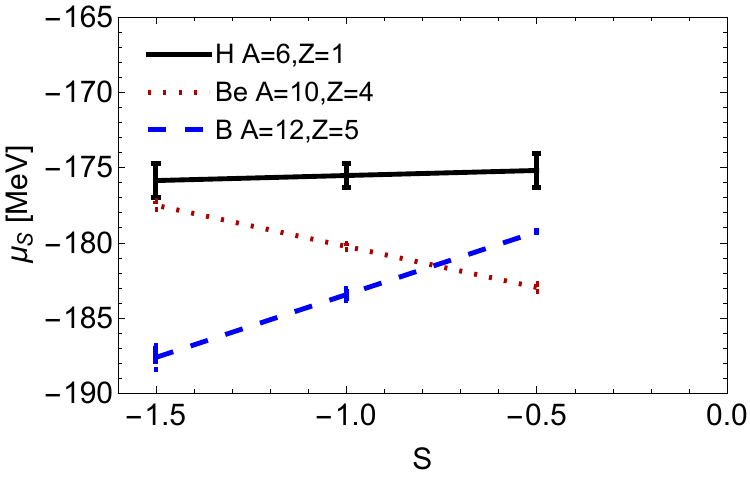}
    \caption{Calculations of the strangeness mesoscopic chemical-potential analog $\mu_S(A,Z)$ versus the strangeness number when 3 mirror (hyper)nuclei are available.}
    \label{fig:etrp0}
\end{figure}

In many EOS applications, particularly when calibrating near isospin-symmetric nuclear matter without explicit strangeness constraints, one effectively works in a macroscopic (thermodynamic) setting where the corresponding chemical potentials satisfy $\mu_Q \simeq 0$ and $\mu_S = 0$ (for example, under conditions of weak equilibrium).
While one may wonder if the large $\mu_S$ value is strongly $S$ dependent such that it approaches zero as $S\rightarrow 0$, the available (hyper)nuclear data shown in Fig.\ \ref{fig:etrp0} do not indicate a strong $S$ dependence within the accessible range. Moreover, $\mu_S\sim 0$ would be far from the simple mass-difference estimate $\mu_S^{EST}\sim -176$ MeV at fixed $A,Z$.
A direct experimental test of the $S\rightarrow 0$ behavior would require additional systems at fixed $A,Z$ with different strangeness content (e.g.\ kaon-bound nuclei, as discussed in the main text).

\begin{table}[]
    \centering
    \begin{tabular}{c|ccccccc}
    \hline
   &   A   & Z & S& $\delta_I$ & $\mu_B$ [MeV] & $\mu_Q$ [MeV] & $\mu_S$ [MeV] \\
   \hline
    Li &   8   & 3 & 0 & $0.3$ & $936$ & $-15.2$ &  \\
     Li &   8   & 3 & -1/2 & $0.19$ &  &  & $-178.14\pm 0.03$ \\
    Li &   8   & 3 & -1 & $0.13$ & $934.8\pm0.1$ & $-6.9\pm0.4$ &  \\
     Be &  9   & 4 & 0 & $0.11$ & $935$ & $-8.7$ &  \\
     Be &   9   & 4 & -1/2 & $0.06$ &  &  & $-177.39\pm 0.07$ \\
    Be &   9  & 4 & -1 & $0$ & $929.2\pm 0.1$ & $-1.6\pm 0.1$ &  \\
         \hline
    \end{tabular}
    \caption{The two points in the phase diagram of $(A,Z)$ where it is possible to directly compare the mesoscopic chemical-potential analogs $\mu_B(A,Z)$ and $\mu_Q(A,Z)$ with and without strangeness. $\mu_S(A,Z)$ is shown for $S=-1/2$ since double-hypernuclei data is unavailable. Note a systematic offset of approximately $\sim -3$ MeV.}
    \label{tab:muBmuSmuQ}
\end{table}

To verify that hypernuclei are self-bound using Eq.\ (\ref{eqn:Gibb_ps}) one ideally requires $\mu_B,\mu_Q,\mu_S$ at the same $\{A,Z,N_\Lambda\}$. However, due to the sparsity of experimental data, we only have two points in the phase diagram where we can obtain $\mu_B,\mu_Q$ at both $S=0$ and $S=-1$.
These calculations actually require 10 nuclei: $(A,Z)$, $(A+1,Z)$, $(A-1,Z)$, $(A,Z+1)$, $(A,Z-1)$ for both $S=0$ and $S=-1$, and the corresponding $\mu_S(A,Z,-1)$ would require double hypernuclei at these specific $A,Z$ combinations ($^{9}_{\Lambda\Lambda}\rm{Be}$ or $^{8}_{\Lambda\Lambda}\rm{Li}$), which is unavailable.
Despite this challenge, we compare the effect of $\mu_B,\mu_Q$ as strangeness is added in Tab.\ \ref{tab:muBmuSmuQ}. We find that $\mu_Q$ decreases as strangeness is added, the additional $\Lambda$ brings the nucleus closer to isospin-symmetric matter, and for $^{9}_{\Lambda}\rm{Be}$ it is precisely at $\delta_I=0$, which gives $\mu_Q\sim 0$.
If we assume that $\mu_S(A,Z,-1)\sim \mu_S(A,Z,-1/2)$, we can estimate our systematic offset of $(p-sT)/n_B$ and we find values of $\sim -3$ MeV, which is consistent with what we found for light nuclei.
Additionally, Tab.\ \ref{tab:muBmuSmuQ} highlights an example of nearly isospin-symmetric matter (i.e.\ $\delta_I=0.056$) where $\mu_S=-177.39\pm 0.07\pm 3$ MeV, illustrating that near-isospin-symmetric nuclei can probe a very large, negative value of $\mu_S$.

\subsection{Remainder for the pressure and entropy}

In Eq.\ (\ref{eqn:Gibb_ps}) we are able to isolate the $(p-sT)/n_B$ contribution given the mesoscopic chemical-potential analogs and $\tilde{E}/A$ extracted from experimental data.
Given that nuclei are self-bound and approximately at vanishing temperatures, we would assume that the left-hand side of the equation should be essentially zero i.e.\ $(p-sT)/n_B\sim 0$ (some very tiny nonzero value may occur if we are not exactly at $T\sim 0$).
In fact, calculating $(p-sT)/n_B$ is a good self-consistency check to understand how well the response of nuclei to changes in $A,Z,S$ can be represented by these mesoscopic chemical-potential analogs.
Furthermore, any corrections in our assumptions such as the system not being exactly at $T=0$ or higher-order Coulomb corrections would show up in a non-zero value of $(p-sT)/n_B$.
Thus, we can consider non-zero values of $(p-sT)/n_B$ as a method to quantify an overall systematic offset.

\begin{figure}
    \centering
    \begin{tabular}{cc}
     \includegraphics[width=0.5\linewidth]{EX/pre_v_A.pdf}    &  \includegraphics[width=0.5\linewidth]{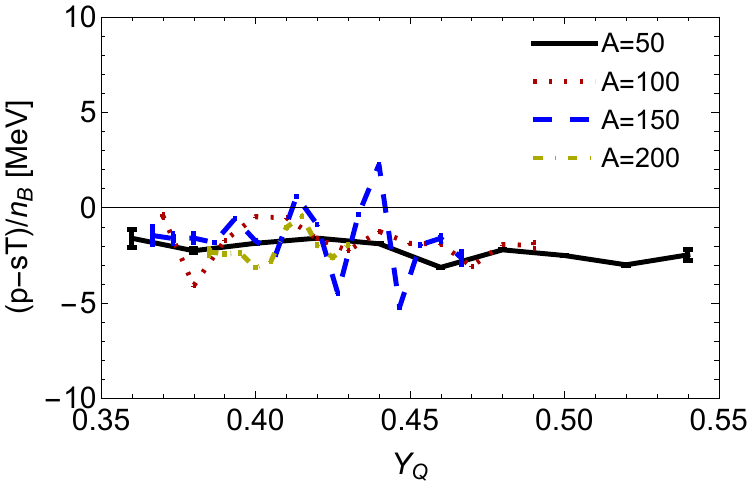}
    \end{tabular}
    \caption{We show our calculated $(p-sT)/n_B$ value from Eq.\ (\ref{eqn:Gibb_ps}) plotted versus $A$ (left) and $Y_Q$ (right). Under our working assumptions (self-bound nuclei and near-zero temperature), this quantity should be consistent with zero within uncertainties. In practice, the extracted values are much smaller than the mesoscopic chemical-potential analogs reported in this work.}
    \label{fig:pressure2}
\end{figure}

In Fig.\ \ref{fig:pressure2} we plot on the left $(p-sT)/n_B$ versus $A$ for different slices of $Y_Q$ and on the right $(p-sT)/n_B$ versus $Y_Q$ for different slices of $A$.
Error bars are calculated using error propagation, assuming entirely uncorrelated errors (such that they are probably overestimated).
We find that there are not large variations in $(p-sT)/n_B$ across our choices in $Y_Q$ and $A$; instead we find a relatively consistent value of $(p-sT)/n_B\sim -2$--$3$ MeV for all our results.

\begin{figure}
    \centering
     \includegraphics[width=0.5\linewidth]{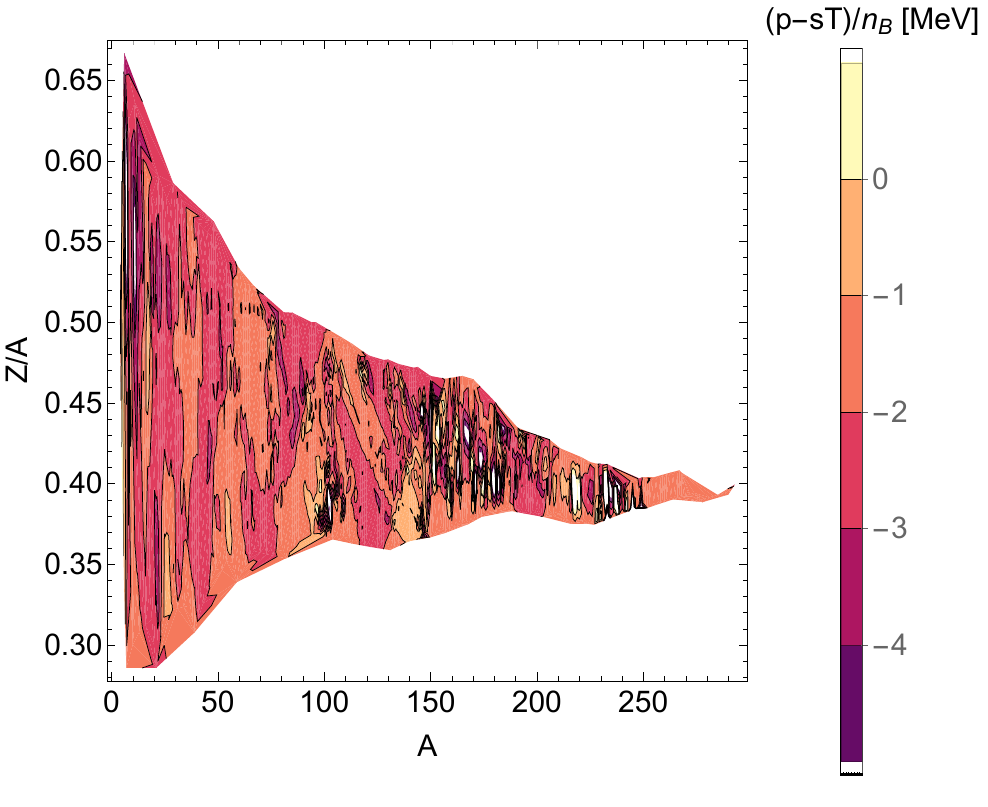}
    \caption{We show our calculated $(p-sT)/n_B$ value from Eq.\ (\ref{eqn:Gibb_ps}) plotted across the nuclear landscape in the $(Z/A,\;A)$ plane. Under our working assumptions (self-bound nuclei and near-zero temperature), this quantity should be consistent with zero within uncertainties. The extracted values are much smaller than the mesoscopic chemical-potential analogs found in this work.}
    \label{fig:pressure3}
\end{figure}
To see this effect more clearly we also plot $(p-sT)/n_B$ across the entire nuclear landscape in Fig.\ \ref{fig:pressure3} (note it is not possible to visualize the error bars in such a plot so only the central values are shown).
We find once again across the entire nuclear landscape similar values that are predominantly negative, on the order of a few MeV.

From our results, it appears we have a very small systematic offset that leads to a negative contribution. Due to the sign and the fact that it does not appear to scale with either $A$ or $Z$, it is unlikely to be due to volume fluctuations or Coulomb effects.
Rather, it may be due to very small finite-$T$ effects that could lead to a non-zero pressure and entropy contribution.
However, we cannot rule out other contributions such as numerical effects or unaccounted-for experimental systematics.
All this being said, we point out that this small shift remains approximately an order of magnitude smaller than $\mu_Q$ and two orders of magnitude smaller than $\mu_B$.
Thus, while it would be interesting to better understand its origin, it does not appear to play a strong role in our results.

\subsection{Other mesoscopic chemical-potential analog results for $\mu_B,\mu_Q,\mu_S$}

\begin{figure}
    \centering
    \includegraphics[width=0.5\linewidth]{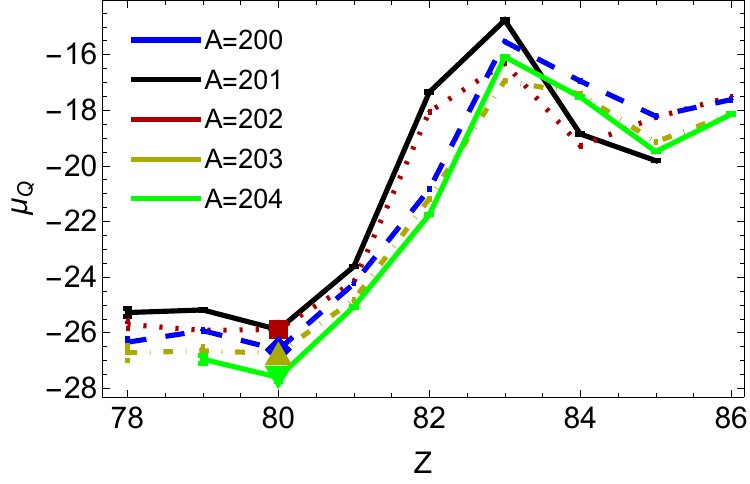}
    \caption{$\mu_Q$ calculations for heavy nuclei ($A=200$--$204$) as they approach the magic number $Z=82$ (peak occurs at $Z=83$). Markers appear for stable nuclei, which occur at the minimum of $\mu_Q$. Here the stable nuclei are all isotopes of Mercury. Results are shown for $\mathcal{O}(h^4)$.}
    \label{fig:magic}
\end{figure}

In Fig.\ \ref{fig:magic} we study the electric-charge mesoscopic chemical-potential analog $\mu_Q$ across a selection group of $A$ slices vs $Z$ that passes across the magic number $Z=82$.
For heavy nuclei, we have found a very robust peak in $\mu_Q$ that occurs as one approaches the magic number of $Z=82$ (lead) from above at $Z=83$ (bismuth), which is one proton above the magic proton shell closure. Since bismuth has one extra proton away from a magic number (and is an unpaired proton), it leads to this peak. We have checked this structure across a wide variety of $A$ values and find that it is robust.

\begin{figure*}
    \centering
    \includegraphics[width=0.45\linewidth]{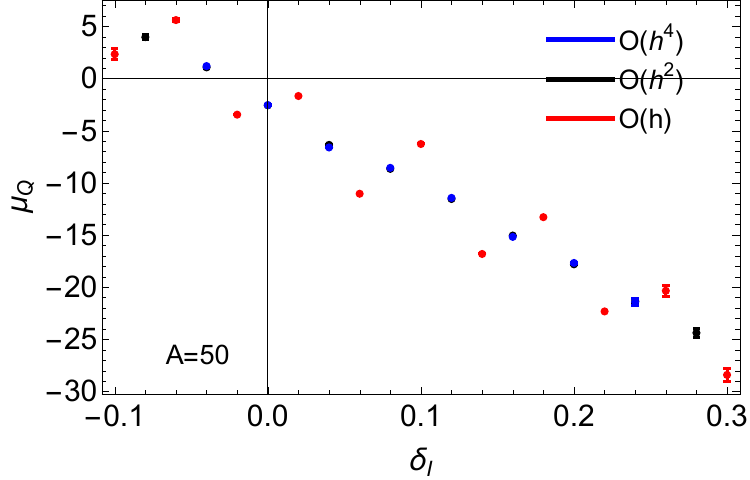}

    \caption{Results for $\mu_Q(A,Z)$ vs the isospin asymmetry. We compare specific chains of isotopes at fixed $A$ and their corresponding error bars. Calculations are performed at different orders of accuracy, using the Euler method $\mathcal{O}(h)$, midpoint method $\mathcal{O}(h^2)$, and $\mathcal{O}(h^4)$.}
    \label{fig:DERVcomp2}
\end{figure*}

In Fig.\ \ref{fig:DERVcomp2} we plot $\mu_Q(A,Z)$ versus the isospin asymmetry for light nuclei, which is the electric-charge mesoscopic chemical-potential analog to Fig.\ \ref{fig:DERVcomp} in the main text.
As with our previous $\mu_B$ results we find that for the Euler derivatives we get fluctuations as protons are added to the system due to pairing effects.
Once again, the $\mathcal{O}(h^2)$ and $\mathcal{O}(h^4)$ results are equivalent.
We note here that we have results at $\delta_I=0$ where $\mu_Q\sim -3$ MeV, which is consistent with the small systematic offset that we estimate using Eq.\ (\ref{eqn:Gibb_ps}).

\begin{figure}
    \centering
    \begin{tabular}{cc}
      \includegraphics[width=0.5\linewidth]{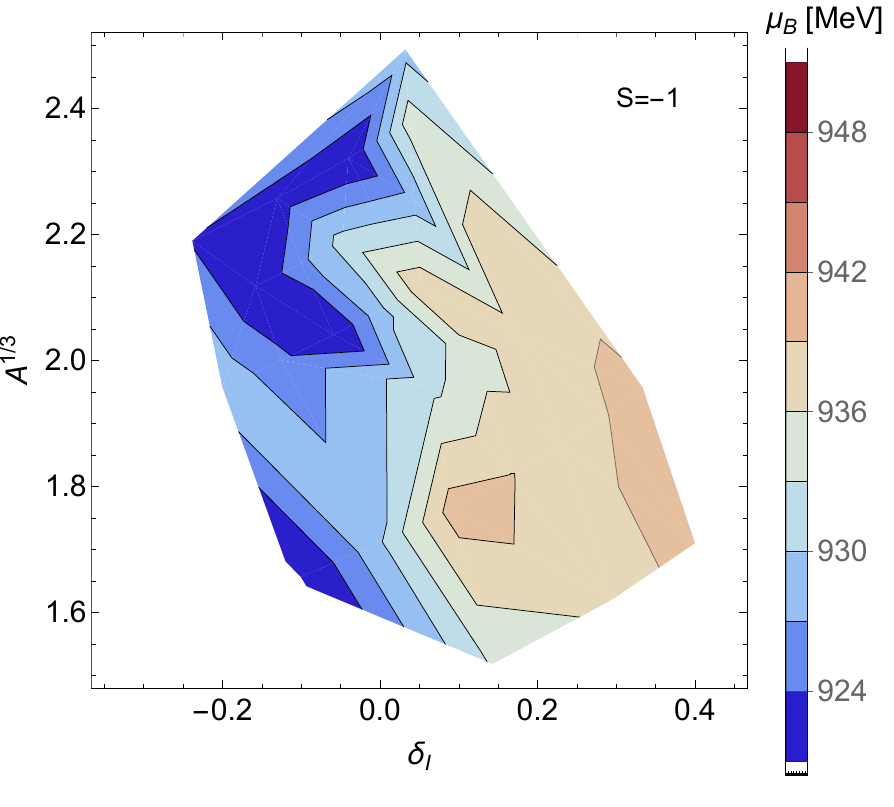}   & \includegraphics[width=0.5\linewidth]{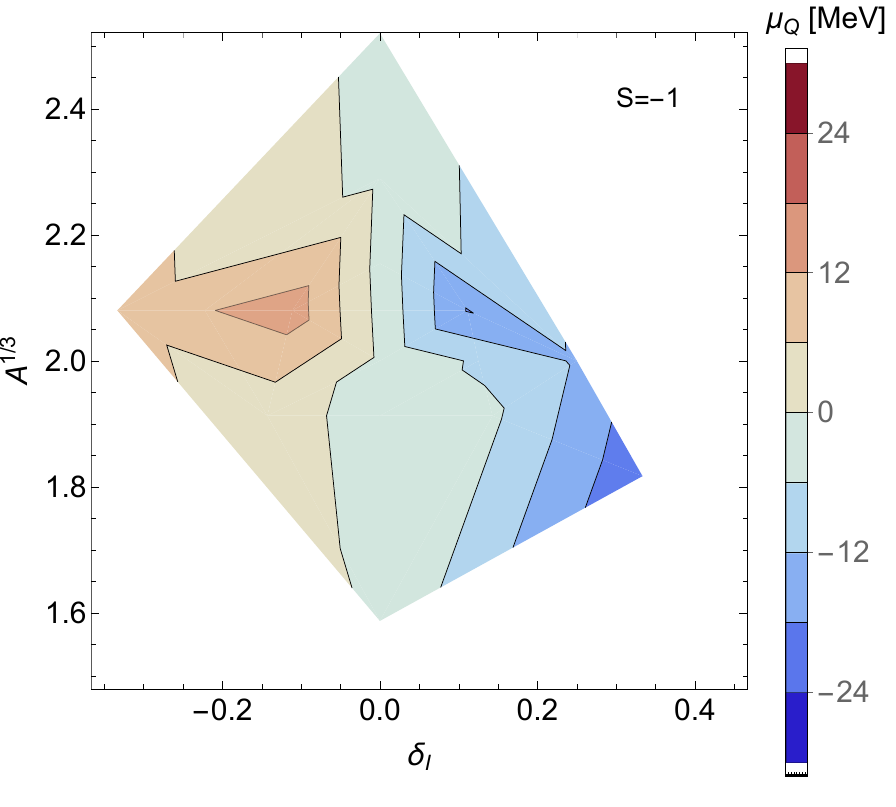}
    \end{tabular}

    \caption{The mesoscopic chemical-potential analogs $\mu_B$ and $\mu_Q$ are shown for hypernuclei, calculated using both the Euler and midpoint methods.}
    \label{fig:muBmuQ_S1}
\end{figure}

In Fig.\ \ref{fig:muBmuQ_S1} we plot the $\mu_B,\mu_Q$ results across the isospin asymmetry $\delta_I$ and volume scaling for hypernuclei states with $S=-1$. These results include both the Euler and midpoint methods within the calculations due to the sparsity of data.
We find comparable results to the results for $S=0$ in that $\mu_B$ is smallest for proton-rich strange matter and largest for neutron-rich strange matter.
Similarly, we find the expected $\mu_Q$ scaling that is positive for $\delta_I<0$ and negative for $\delta_I>0$.
We note that since only $\Lambda$ baryons are within the hypernuclei we do not anticipate strong differences compared to the light nuclei.
However, other types of hyperons would likely have strong effects since some carry electric charge as well.

\end{document}